\documentstyle[twocolumn,aps]{revtex}

\begin{document}

\title{Scar and Antiscar Quantum Effects in Open Chaotic Systems}
\author{L. Kaplan\thanks{kaplan@physics.harvard.edu}
\\Department of Physics and Society of Fellows,\\ Harvard
University, Cambridge, MA 02138}
\maketitle

%\centerline{\bf DRAFT: DO NOT DISTRIBUTE}

\begin{abstract}

We predict and numerically observe strong periodic orbit effects in the
properties of open quantum systems with a
chaotic classical limit. Antiscars lead to a large number of exponentially
narrow resonances when the opening is located on a short unstable orbit
of the closed system; the probability to 
remain in the system at long
times is thus exponentially enhanced over the random matrix theory
prediction. The distribution of resonance widths and the probability to
remain are quantitatively given in terms of only the stability
matrix of the orbit on which the opening is placed. The long-time
remaining probability density is non-trivially distributed over the
available phase space; it can be enhanced or suppressed near orbits other than
the one on which the lead is located, depending on the periods and classical
actions of these other orbits. These effects of the short periodic orbits on
quantum
decay rates have no classical counterpart, and first appear on times scales
much larger than the Heisenberg time of the system.
All the predictions are quantitatively compared
with numerical data.

\end{abstract}

\section{Introduction}

Wavefunction scarring, the enhancement or suppression of quantum
eigenstate intensity along an unstable orbit of the corresponding classical
system, is a fascinating and generic property of quantum chaotic behavior.
Along with dynamical localization, it is one of the striking ways in which
a quantum system can show deviation from ergodicity at the single-channel
level even though the classical dynamics is completely ergodic. Wavefunction
intensities near a short unstable periodic orbit follow a distribution
far from that predicted by random matrix theory (RMT), with some wavefunctions
having much more intensity and other much less than would be predicted based
on gaussian random fluctuations. The phenomenon is at first glance paradoxical,
because the long-time (and indeed stationary) quantum behavior retains a memory
of the short-time classical motion, a memory that is completely absent in the
long time {\it classical} dynamics of a chaotic system. Scarring has been
observed experimentally in a wide variety of systems, including
microwave cavities\cite{sridhar,stockman}, semiconductor
structures\cite{fromhold},
and the hydrogen atom in a magnetic field\cite{wintgen,wintscar}.

A theory of scarring based on the linearized evolution of gaussian
wavepackets was first provided in~\cite{hellerscar}; later theoretical
work by 
Bogomolny\cite{bogomolny} in coordinate space and
Berry\cite{berry} in Wigner phase space 
followed. These made predictions about the average
intensity on a classical periodic orbit of states in a given energy band;
however, because of the energy smoothing involved, no predictions were
possible about the statistical properties of individual peak heights in the
local density of states. More recently a nonlinear theory was
developed~\cite{nlscar} which made it possible to predict the statistical
properties of individual wavefunctions, in the semiclassical
limit. A homoclinic orbit analysis showed that long-time return amplitude
to the vicinity of a periodic orbit bore the imprint of the short-time
linearized classical dynamics around the periodic orbit. This leads
to a natural separation of scarring intensity into a classical
short-time component and a random long-time component, as suggested
already in\cite{ott}.

In\cite{sscar}, predictions were made about the distribution of wavefunction
intensities on a periodic orbit and at a generic point in phase space.
The full distribution of intensities, which includes samples taken over
all of phase space, has a long tail (compared to the Porter-Thomas prediction
of RMT), dominated by the effect of the least unstable periodic orbit.
The functional form of this tail is given in the semiclassical (high-energy)
limit very simply in terms of the stability exponent of this least
unstable orbit, as long as an optimally-oriented test basis is chosen. 
Furthermore, upon ensemble averaging a power-law intensity distribution tail
is obtained, in sharp contrast with the exponential tail predicted by RMT.
This result is also to be contrasted with the log-normal intensity
distribution tail which obtains in diffusive systems~\cite{lognormal,lupusax}.
Thus, although RMT is accepted as the zeroth-order approximation
for both chaotic and disordered quantum systems, deviations from RMT
predictions can be qualitatively different in the two cases, providing
an impetus for the present research.

The numerically tested quantitative predictions in ~\cite{nlscar,sscar}
concerned the local densities of states of closed systems. Experimentally,
certain properties of open systems, such as resonance widths and conductances
may be more amenable to experimental verification~\cite{openexper}. Much
work has been done here within the context of RMT~\cite{rmt}; deviations
from RMT in disordered systems have also been considered\cite{disorder}. In the
present work, we address the distribution of decay lifetimes in a leaky
chaotic system, and the probability to remain in such a system\cite{miller}
as a function
of time and the location of the ``leak" (open channel). The distribution
in phase space of the remaining probability density at long times is
considered, as well as the dependence of the probability to remain on
where the particle was first ``injected" into the system. These last two
questions bring us into contact with wavefunction correlations and
transport in chaotic systems, which (as we show) can
be very different from RMT
expectations where periodic orbits are involved. All the quantitative
predictions which follow are tested numerically.
A study of conductance 
properties in two-lead chaotic systems, including mean conductance,
conductance fluctuations, distribution of peak heights,
and peak correlations (and how all these depend on the placement
of one or both leads in relation to the classical orbits) is
forthcoming~\cite{bies}.

\section{Classical and quantum weakly open chaotic systems}

\subsection{Classical behavior}

We begin by considering a small opening in a classically chaotic system,
which allows a particle to escape from the system. We will often
use language suggesting that the ``opening" is defined in position
space, as it would often be, for example, in a mesoscopic experiment.
However, the formalism considered here is much more general: all
that is required is that the opening be localized in the classical
{\it phase space}; escape routes that are defined exclusively in terms
of position or momentum are special cases of this. A simple example of a
momentum space opening is a potential barrier that allows particles to leave
only if their momentum is directed almost normal to the wall\cite{stone}.
An opening having the shape of a phase-space gaussian naturally
occurs when one considers tunneling out of a metastable chaotic well 
formed by a continuous potential~\cite{creagh}.
Now we can imagine
forming a mesh in classical phase space with each cell the size of the
opening; because the classical dynamics (in the closed system) is chaotic, 
probability density starting in one such cell will soon be evenly
distributed over all the available cells. The time for this to happen
is logarithmic in the size of the opening $w$:
\begin{equation}
T_{\rm mix} \sim {1 \over \overline\lambda} |\log w| \,.
\end{equation}
Here $\overline\lambda$ is the Lyapunov exponent of the classical dynamics
(the mean rate of chaotic divergence of classical orbits),
and the total size of
phase space, in terms of which $w$ must be measured, has been set to unity.
On the other hand, the escape time from the system is inversely
proportional to the leak size $w$, so a small value of $w$ will cause
complete mixing of the remaining probability
to take place on a time scale much shorter than the scale on which 
probability is leaking out. One obvious consequence is that the probability
to remain in the open classical system follows an exponential law. This
behavior is, of
course, independent of the position of the leak. Also,
density is constantly redistributing itself, so that the remaining probability
density remains evenly distributed on the scale of our mesh, except
in a corridor of length scaling as $1/\overline\lambda$ leading forward
in time from the position of the opening. Notice that the width of this
corridor where the probability to remain is suppressed scales as the size
of the opening. Thus, this corridor has no effect on the quantum behavior
when the opening is small compared to $\hbar$.
 Finally, even if the
initial probability is not evenly distributed over the entire phase space,
the long-time behavior is unaffected (as long as the bulk of the
probability is not initially placed in a corridor similar to the
one described above, but leading {\it backwards} in time from the opening).

In contrast to these results, we will find in the quantum case that the 
probability to remain in the system at long times depends strongly
on whether the opening is located on a classical (unstable)
periodic orbit, even though
the initial probability density is evenly distributed. Again, we see
that long-time quantum behavior retains a better memory of short-time
classical dynamics than does the long-time classical behavior.
Also, we will see that given a leak placed on a periodic orbit,
the remaining probability distribution at long times can be strongly affected 
not just on the periodic orbit itself, but also on the {\it other}
short periodic orbits of the system. Enhancement or suppression can be observed
depending on the energy range considered and on the classical actions of 
the orbits in question. Similarly, the probability to remain at long times
will be affected if the original probability is injected on an unstable
periodic orbit different from the one where the opening is located.
All this is true even though the decay is taking place on a time scale
much longer than any other time scale in the problem (the period of the
short orbit, the mixing time, and also
the Heisenberg time, i.e., the inverse of the mean level spacing).

\subsection{Quantum mechanics and RMT}

Let the quantization of our classical system be given by an $N-$dimensional
Hilbert space ($N$ is the number of Planck-sized cells in the classical
phase space), with dynamics in the closed system
given by the Hamiltonian $H_0$. If the
opening is very small (less than one open channel, so that the resonances
are non-overlapping), we can write an effective Hamiltonian for the open
system:
\begin{equation}
\label{hdef}
H=H_0 -i{\Gamma \over 2} |a\rangle \langle a| \,,
\end{equation}
where $|a\rangle$ is a quantum channel associated with the opening, and
$\Gamma$ is the decay rate in that channel (taken to be small).
$|a \rangle$ could be a gaussian wavepacket enclosing the hole, or a position
or momentum state. It is important to note here that the opening
is small compared to $\hbar$ (less than half a wavelength if in position
space). One can of course consider large openings, or ones which are not
thus localized to a single channels; these possibilities are considered
towards the end of the present paper. We emphasize, however,
that the phenomenon 
discussed here is already present in its full form for the tiniest
single opening, without the complications that arise in the more general case.

For a small hole, the main effect of the opening on a wavefunction
$|\Psi_n\rangle$ of the closed system is that it acquires a decay width
proportional to the intensity of the wavefunction at the opening:
\begin{equation}
\Gamma_n = \Gamma |\langle \Psi_n|a\rangle|^2 \,.
\end{equation}
If the intensities $x_n \equiv N|\langle \Psi_n|a\rangle|^2$
follow a chi-squared distribution, as in RMT, we have probability
$P(x)={1 \over \sqrt{2\pi x}} \exp{(-x/2)}$ for real overlaps
$\langle \Psi_n|a\rangle$, and 
$P(x)=\exp{(-x)}$ for complex overlaps. Consider the complex case.
Because mixing between the states of the closed system can be neglected
in the small $\Gamma$ regime, the total probability to remain in the system
is given by a sum over these states:
\begin{eqnarray}
P_{\rm rem}(t) & = & {1 \over N} \sum_{n=0}^{N-1} e^{-{x_n \over N} \Gamma t} 
\nonumber \\
& = & \int_0^\infty dx P(x) e^{-x \Gamma t/N} \nonumber \\
& = & \int_0^\infty dx e^{-x} e^{-x \Gamma t/N} \nonumber \\
& = & {1 \over 1+ \Gamma t/N} \,.
\end{eqnarray}
[Remember that $N$ is the total number of states in the system; the classical
decay rate is given by $\Gamma_{\rm cl} = \Gamma/N$
because only one channel has the possibility
to decay.] We see that at short times ($t \ll \Gamma_{\rm cl}^{-1}$), the
probability to remain in the system is $P_{\rm rem}(t)
\approx 1-\Gamma_{\rm cl} t$,
as expected, while at long times we have the asymptotic behavior
\begin{equation}
P_{\rm rem}(t) \approx { 1 \over \Gamma_{\rm cl} t} \,.
\end{equation}
In the case of $M$ independent weakly open channels, i.e.
\begin{equation}
\label{hmany}
H=H_0 -i\sum_{i=0}^{M-1} {\Gamma^{(i)} \over 2} |a^{(i)}\rangle \langle 
a^{(i)}| \,,
\end{equation}
the classical decay rate is given by 
\begin{equation}
\Gamma_{\rm cl} = {1 \over N} \sum_{i=0}^{M-1} \Gamma^{(i)} \,,
\end{equation}
and the RMT probability to remain is 
\begin{equation}
\label{pmany}
P(t) = \prod_{i=0}^{M-1} {1 \over 1+ \Gamma^{(i)} t/N} \,.
\end{equation}
Taking $M \to \infty$ while keeping the total decay rate $\Gamma_{\rm cl}$
constant, exponential decay consistent with the classical prediction is
obtained. On the other hand, fixing the number of channels $M$ and taking
$t \to \infty$, we observe the power-law behavior
\begin{equation}
P_{\rm rem}(t)= { (N/t)^M \over \prod_{i=0}^{M-1} \Gamma^{(i)} }
\end{equation}

The case of real overlaps $\langle \Psi_n|a\rangle$ follows similarly:
each real random overlap counts as half of a complex one, so there
$P_{\rm rem}(t) \sim t^{-M/2}$.
In the literature one often considers the distribution
of delay times
for scattering off of the system in question: there
one must include
the probability of populating a given resonance in the first place,
which of course is
proportional to $\Gamma_n$. This leads to an extra factor of $t$ 
in the denominator, giving $P_{\rm delay}(t) \sim t^{-M-1}$
for complex overlaps
and $P_{\rm delay}(t) \sim t^{-M/2-1}$ for real overlaps.
In our case, we imagine
the system to be populated first, before the lead is opened up, and thus no
extra power of $t$ is present.

\section{Effect of periodic orbits}

\subsection{Probability to remain}

We now go beyond RMT to consider the effect of real dynamics on the 
quantum probability to remain in a classically chaotic system. Take the channel
$|a\rangle$ to be on or near an (unstable)
periodic orbit of instability exponent $\lambda$. The smoothed
local density of states
at $|a\rangle$ is obtained by fourier transforming its short-time
autocorrelation function, which is easily obtained by linearizing the
classical equations of motion near the unstable
orbit~\cite{hellerscar,nlscar,sscar}.
Thus, for example, if the periodic orbit in question is a fixed point
of a discrete time map, and $|a\rangle$ is a gaussian wavepacket optimally
aligned along the stable and unstable manifolds of the orbit, the
short-time autocorrelation function is given by
\begin{equation}
\label{alin}
A_{\rm lin}(t) \equiv
 \langle a | a(t) \rangle = {e^{-i \phi t} \over \sqrt{\cosh{\lambda t}}}
\end{equation}
Here $-\phi$ is a phase associated with one iteration of the orbit:
it is given by the classical action in units of $\hbar$, plus a Maslov
phase as appropriate. The subscript ``lin" indicates that the
expression is obtained within the linearized classical approximation; 
it is valid on time scales short compared to the mixing time
$T_{\rm mix} \sim {1 \over \overline\lambda} |\log \hbar|$.

A more general form of Eq.~\ref{alin} applies for a non-optimally 
oriented wavepacket (e.g. a position or momentum channel could be non-optimal
depending on the direction of the invariant manifolds at the periodic point),
and also for a channel not exactly centered on a periodic point
\cite{bies,genforms}.
In particular, for a wavepacket centered on the periodic orbit but not
optimally oriented with respect to its invariant manifolds, the form above
becomes
\begin{equation}
\label{alingen}
A_{\rm lin}(t) =
 \langle a | a(t) \rangle = {e^{-i \phi t} \over \sqrt{\cosh{\lambda t}
+iQ\sinh{\lambda t}}} \,.
\end{equation}
In Eq.~\ref{alingen}, $Q$ is a non-optimality parameter: in a
coordinate system where the stable and unstable manifolds are orthogonal, $Q$
is a function of the angle between the orientation of the phase-space gaussian
(at some fixed eccentricity)
and either of these two directions. Alternatively, if the wavepacket
$|a\rangle$ is fixed to have a circular shape in phase space (i.e. to have
equal
and uncorrelated uncertainties in $q$ and $p$), $Q$ becomes a function of
the non-orthogonality between the stable and unstable manifolds.
In any case, as long as $Q$ is not very large,
the qualitative behavior is not much changed,
and although analytic results are less easy to obtain for non-zero $Q$,
quantitative predictions can be readily produced for comparison
with any experimental or numerical data.

The key point for our purposes here is that for a small exponent
$\lambda$, the autocorrelation function remains large for the first
$O(\lambda^{-1})$ iterations of the orbit, and the local density of states
has a short-time envelope
\begin{equation}
S_{\rm lin}(E) \equiv \sum_t e^{i E t} A_{\rm lin}(t)
\end{equation}
of width scaling as $\lambda$ and height scaling as $\lambda^{-1}$ [also see
Fig.~\ref{figslin} below].

Nonlinear recurrences on time scales beyond the mixing time (associated
with orbits homoclinic to the original periodic orbit) lead to fluctuations
multiplying this  spectral envelope, eventually producing a line spectrum
\begin{equation}
S(E) = \sum_n |\langle \Psi_n|a\rangle|^2 \delta(E-E_n) \,.
\end{equation}
The line heights $x_n=N|\langle \Psi_n|a\rangle|^2$ are distributed in each
energy region as
a chi-squared distribution with mean $S_{\rm lin}(E)$ \cite{sscar}:
\begin{equation}
P(x)={1 \over S_{\rm lin}(E)} e^{-x/S_{\rm lin}(E)} \,.
\end{equation}
Thus, the distribution of decay widths can be strongly energy-dependent;
in particular, the probability to remain in the system at long times
is now given by
\begin{eqnarray}
\label{preme}
P_{\rm rem}(t) & = & \int_0^\infty dx P(x) e^{-x \Gamma t/N}  \nonumber \\  &=& 
{1 \over 1 + S_{\rm lin}(E) \Gamma t/N} \nonumber \\ & \to &
{1 \over S_{\rm lin}(E)}  { 1 \over \Gamma_{\rm cl}t}
\end{eqnarray} 
if initially only states with energy around $E$ are populated.

The scarred
states (those with energy close to satisfying the EBK quantization
condition) have $S_{\rm lin}(E)>1$ and thus decay much faster than the
antiscarred states~\cite{ott},
which are far from satisfying EBK and thus have
$S_{\rm lin}(E)<1$. Let us examine more closely these two distinct
energy regimes.
Near the quantization energy $E=\phi$, the smoothed density of states
$S_{\rm lin}(E)$ has its peak; its height scales inversely with
$\lambda$ for small $\lambda$~\cite{hellerscar}:
\begin{equation}
\label{maxs}
S_{\rm lin}(E=\phi) \approx c/\lambda \,,
\end{equation}
where $c=5.24$ is a numerical constant~\cite{sscar}. The width
\footnote{
We note that our presentation here is in the context of  a
discrete-time map; thus $E$ is a dimensionless quasienergy
that takes values in the interval $[0,2\pi]$. For a real
continuous-time system with a periodic orbit of period $T_P$,
it is of course the quantity $ET_P/\hbar$ that must be compared with the
dimensionless number $\lambda$. Also, the smoothed density of states
will then have an infinite sequence of peaks, each centered on an energy
satisfying the EBK quantization condition\cite{nlscar}. The ratio of
each peak width to the spacing between peaks scales as $\lambda$.}
of this peak in $S_{\rm lin}(E)$
scales linearly with $\lambda$ for small $\lambda$, and all of the anomalously
enhanced wavefunction intensities come from this energy region, as was
observed and confirmed numerically in \cite{sscar}. In the open system, these
states produce an excess of large resonance decay widths and decay faster (by
a factor of $O(\lambda^{-1})$) than would be predicted by RMT.

Because our focus here is on the long-time behavior of weakly open systems,
we are more interested in the (complementary) suppression of the
smoothed local density of states far from the resonance energy. Again, we
consider the strong scarring regime, where $\lambda \ll 1$: then the
linear spectrum falls off exponentially far away from the peak,
\begin{equation}
\label{inters}
S_{\rm lin}(E) \approx {2 \pi \over \lambda} e^{-\pi |E-\phi| /2 \lambda}
\end{equation}
for $|E-\phi| \gg \lambda$. Within $O(\lambda)$ of the optimal
anti-scarring energy,
$E=\phi+\pi$, the spectrum deviates from the exponential law 
and smoothly approaches the value
\begin{equation}
\label{mins}
S_{\rm lin}(E=\phi+\pi) \approx {4 \pi \over \lambda} e^{-\pi^2/2 \lambda}
\end{equation}
at the minimum. The region within $O(\lambda)$ of $E=\phi+\pi$ is thus
responsible for producing the smallest wavefunction intensities, and
the narrowest resonances in the corresponding open system. This excess of
exponentially small decay rates is as dramatic a signature of the
underlying classical behavior as the long wavefunction intensity tails found
in \cite{sscar}. As we will observe in the next section, the antiscarring
effect on the long-time behavior of open systems can be very striking even
for moderate exponents
$\lambda$ (e.g. $\lambda \approx 1$), as long as the lead is
optimally placed with respect to the periodic orbit.

Smoothed local densities of states $S_{\rm lin}(E)$ on periodic orbits
of instability exponents $\lambda=0.20$ (solid curve)
and $\lambda=0.15$ (dashed curve)
are  shown graphically in 
Fig.~\ref{figslin}. The figure can be viewed as representing the average
wavefunction intensity in a closed system as a function of energy, or
the mean resonance width (in units of $\Gamma_{\rm cl}$)
at that energy in the weakly open system.
The phase $\phi=0$ has been chosen
so as to make $E=0$ the EBK energy at which maximum scarring occurs
(Eq.~\ref{maxs}).
A half-log scale is used to emphasize the exponential falloff
in average resonance width between $E=0$ and $E=\pi$ (Eq.~\ref{inters}),
and the minimum
near the anti-EBK energy $E=\pi$ (Eq.~\ref{mins}). For reference, the smoothed
local density of states in RMT (applicable when the lead
is not in the vicinity of any short periodic orbit) is displayed as a dotted
line in the figure.

We now consider the energy-averaged probability to remain in the open system:
this will be the quantity studied in detail numerically in the next section,
where the model system is a (non-energy conserving) discrete-time kicked map.
[For an energy-conserving system, varying the strength of a weak
magnetic field and thus sweeping through different values of the phase
$\phi$ would produce the same result.]
Again, because the perturbation induced by opening up the system is small,
there is little mixing among states of different energy. Thus the total
probability to remain is obtained simply by averaging the probabilities
at the different energies. From Eq.~\ref{preme} we see that at short times,
the classical behavior is recovered:
\begin{equation}
P_{\rm rem}= 1- <S_{\rm lin}> \Gamma t/N = 1- \Gamma_{\rm cl}t \,,
\end{equation}
as $<S_{\rm lin}>=A_{\rm lin}(0)=\langle a|a\rangle =1$ by normalization.
Thus at short times, $t \ll \Gamma_{\rm cl}^{-1}$, the faster-decaying
scarred states and slower-decaying antiscarred states always cancel exactly
and no sign of the quantum signatures of the
underlying classical dynamics can be observed.
On the other
hand, at long times, i.e. for
$t \gg [S_{\rm lin}(E_{\rm min})\Gamma_{\rm cl}]^{-1}$, we obtain the
very different behavior
\begin{equation}
P_{\rm rem}(t)={<S_{\rm lin}^{-1}> \over \Gamma_{\rm cl} t} \,.
\end{equation}
Here $E_{\rm min}$ is the energy at which the smoothed spectrum has its
minimum; for an optimally placed lead $|a\rangle$
this energy is exactly $\pi$ out of phase with the EBK energy $\phi$, as
discussed above (Eq.~\ref{mins}). $<S_{\rm lin}^{-1}>$ is the
inverse of the smoothed density of states at $|a\rangle$, averaged
over energy (or weak magnetic field).

As $<S_{\rm lin}>=1$ by definition, any fluctuations
in the smoothed spectrum resulting from short-time recurrences will
cause $<S_{\rm lin}^{-1}>$ to be greater than one, resulting in
an enhanced probability to remain at long times. In particular, for
an optimally placed lead (corresponding to Eq.~\ref{alin}), let us consider
the strong scarring regime of small $\lambda$. This gives the exponentially
large enhancement
\begin{equation}
\label{enh}
<S_{\rm lin}^{-1}> = \left({\lambda \over 2 \pi} \right)^{2}
e^{\pi^2/2 \lambda} \,.
\end{equation}
This long-time behavior is completely dominated by the most antiscarred
states, i.e. those with energy within $O(\lambda)$ of $E_{\rm min}=\phi+\pi$
(Eq.~\ref{mins}).
For a non-optimally placed lead, with not too large 
non-optimality parameter $Q$
(see Eq.~\ref{alingen} and discussion following) we find empirically
a similar exponential enhancement of the typical decay time:
\begin{equation}
<S_{\rm lin}^{-1}> = \left({\lambda \over 2 \pi} \right)^{2}
e^{(\pi^2/2 -bQ)/\lambda} \,,
\end{equation}
where $b=1.1$ is a numerical constant.

If the state $|a\rangle$ defining the phase-space location of the opening 
is centered off of the periodic orbit, but within $\hbar$ of the orbit,
one still has
fluctuations in the linear density of states and consequently an enhancement
in the probability to remain at long times. An analytic form for the
linear autocorrelation function in such a case can be found in\cite{genforms}.
For a circular minimum-uncertainty phase-space opening centered a distance
$\delta$ away from a periodic orbit with small exponent $\lambda$,
the energy-averaged value $<S_{\rm lin}^{-1}>$ scales as
\begin{equation}
\label{deltaeq}
<S_{\rm lin}^{-1}> \sim \lambda^2 e^{(\pi^2/2-d\delta/\sqrt\hbar)/\lambda} \,,
\end{equation}
where $d$ is yet another numerical constant. $\delta$ can be a displacement
along either the stable or unstable direction away from the orbit.
Thus, deviations from
RMT behavior are observed in an area scaling as $\hbar$ surrounding
the periodic orbit. Maximum enhancement of $S_{\rm lin}^{-1}$ (i.e.
enhancement of order $\lambda^2 e^{\pi^2/2\lambda}$)
occurs
for $\delta < O(\lambda \sqrt \hbar)$, corresponding to a phase-space
area scaling as $\lambda^2 \hbar$ surrounding the orbit. Thus, if we consider
the long-time probability to remain in the system {\it averaged over all
possible
positions of the lead}, we obtain
\begin{equation}
P_{\rm rem}  = { 1+ O(\hbar \lambda^4 e^{\pi^2/2\lambda}) \over 
\Gamma_{\rm cl}t} \,.
\end{equation}
[The correction to RMT is obtained by multiplying the maximum
obtainable enhancement by the size of the phase-space region where such
enhancement occurs.] In principle, contributions from all the periodic
orbits need to be added, however, if orbits with small $\lambda$
exist, they will clearly dominate any such sum. The result is that
at finite energy, exponentially large (in $1/\lambda$) deviations from
RMT are found even in the phase-space averaged analysis. In the
$\hbar \to 0$ limit of any given classical system, the RMT
behavior is recovered because the chance of a lead being found on the
short periodic orbit goes to zero.

In Section~\ref{numprob}, we present theoretical predictions and numerical
data measuring the probability to remain in the system at long times as
a function of the location of the opening.

\subsection{Probability density at long times and dependence on initial
conditions}

\label{dens}

Up until now we have been focusing on the distribution of resonance widths
and on the {\it total}
probability to remain in the system starting from a {\it uniform}
initial state, all as a function of the location of the lead $|a\rangle$.
In other words, while changing the location of the opening, we have always
been tracing over the initial and final states of the system. We now proceed
to address two related questions, both of which require us to consider
transport properties and wavefunction intensity correlations within the system.

First, still taking the initial filling to be uniform, and fixing the location 
of the opening to be on a periodic orbit, it is natural to
ask how the long-time probability density remaining in the system distributes
itself over phase space. Classically, we expect the remaining probability
always to be redistributing itself on a time scale short compared to the
decay time, and thus to be uniform except in a very narrow corridor
encompassing the unstable orbit. The width of the corridor scales as
the size of the opening.
In RMT, of course, the remaining probability is
also completely uniform except at phase space locations having non-zero
overlap with $|a\rangle$. In contrast, we find that in the real quantum
system, the remaining probability density is strongly suppressed in 
a corridor of size $\hbar$ around the orbit,
much wider than the size of the lead. Even more
interesting is the fact that the probability density can be either
relatively enhanced or suppressed along the {\it other} unstable orbits
of the system, depending on the classical actions associated with these
orbits.

Before proceeding, we mention a closely related problem, which can be thought
of as a time-reversed version of the one stated above. Instead of initially
filling the system with a uniform density, we inject probability in some
known initial state and look at the probability to remain after a long time
as a function of this initial state. This state, which
we call $|b\rangle$, should be classically
well-defined, i.e. it can be a phase-space gaussian, or a position
or momentum state, as discussed above. 

The two problems are in general distinct: if $H$ is the non-Hermitian
quantum Hamiltonian, the first involves the quantity
$\langle b |e^{-iH^\dagger t}e^{iHt}|b \rangle$, while the second measures
$\langle b |e^{iH^\dagger t}e^{-iHt}|b \rangle$. However, when $\Gamma$ is
very small (in the regime of non-overlapping resonances), $H$ is nearly
normal, the distinction between left and right eigenstates vanishes,
and the two quantities both converge to the eigenstate sum
\begin{equation}
\label{premb}
P_{\rm rem}^b(t) = \sum_n |\langle b|\Psi_n\rangle|^2 e^{-\Gamma_n t} 
\end{equation}
For $|b\rangle$ not on the periodic orbit containing the lead $|a\rangle$,
the quantity $|\langle b|\Psi_n\rangle|^2$ is independent
of $\Gamma_n \sim |\langle a|\Psi_n\rangle|^2$, and
follows its own
chi-squared distribution with mean scaling as $S_{\rm lin}^b(E)$.
Here $S_{\rm lin}^b(E)$ is the fourier transform of the
linearized (short-time) autocorrelation function of the test state $|b\rangle$;
it is to be distinguished from $S_{\rm lin}(E) \equiv S_{\rm lin}^a(E)$, the
smoothed local density of states {\it at the lead}. We easily obtain,
at energy $E$,
\begin{eqnarray}
P_{\rm rem}^b & = &
{S_{\rm lin}^b(E) \over 1 + S^a_{\rm lin}(E) \Gamma t/N} \nonumber \\ & \to &
{S_{\rm lin}^b(E) \over S^a_{\rm lin}(E)}  { 1 \over \Gamma_{\rm cl}t} \,.
\end{eqnarray} 

Averaging over $E$, we obtain the ratio of the remaining probability density
at $|b\rangle$ to the average remaining density at long times:
\begin{equation}
\label{ratio}
{ P_{\rm rem}^b \over P_{\rm rem} } =  {< S_{\rm lin}^b / S_{\rm lin}^a >
\over  < 1 /S_{\rm lin}^a > }\,.
\end{equation}
We see that this ratio goes to unity if $S_{\rm lin}^b$ has no energy
dependence, i.e. if $|b\rangle$ does not lie on a short periodic orbit.
If the position $|a\rangle$ of the lead itself does not lie on a periodic
orbit, the remaining density profile will of course be flat
over {\it all} states $|b\rangle$. However, if both $|a\rangle$ 
and $|b\rangle$ lie on periodic orbits, the probability to be found at
$|b\rangle$ can be either suppressed or enhanced, depending on
whether the energy envelopes $S_{\rm lin}^a$ and $S_{\rm lin}^b$ are
in or out of phase in the energy range being averaged over. For simplicity,
let us consider an example where the periods, and instability exponents
of the two orbits are equal. Then the two smoothed energy envelopes are
identical, up to a relative phase shift (the difference between 
$\phi_a$ and $\phi_b$), which can be adjusted by varying a magnetic flux
enclosed by one of the orbits. If the two are exactly in phase, 
$S_{\rm lin}^a=S_{\rm lin}^b$, then the ratio in Eq.~\ref{ratio} reduces
to $1/<S_{\rm lin}^{-1}>$, which, we recall from our previous discussion,
is a quantity exponentially small in the instability exponent
$\lambda$. Thus, the remaining probability very strongly avoids the orbit
on which $|b\rangle$ is located. Another way of expressing this result
is that the total probability to remain in the system at long times is
exponentially suppressed if the initial state is located on an orbit
which is ``in phase" with the orbit on which the opening is located.

The suppression of probability density given by
Eq.~\ref{ratio}
is of course a pure quantum interference phenomenon; it has no
analogue in the classical dynamics of open systems.
It is also fundamentally a long-time effect as there
is in general no short path leading from  $|a\rangle$ to $|b\rangle$
which could give rise to such intensity correlations.
However, despite being intrinsically long-time and quantum, the phenomenon
can be understood only in terms of the {\it short-time, classical}
dynamics near each of 
the two unstable periodic orbits. This demonstrates once again
the power of semiclassical techniques for understanding long-time
quantum behavior.

In the opposite extreme case, where the two orbits are out of phase
exactly by $\pi$ [$S^b_{\rm lin}(E)=S^a_{\rm lin}(E+\pi)$],
the ratio in Eq.~\ref{ratio} is dominated by the
region of the envelope where $S^b_{\rm lin}$ is maximized and
$S^a_{\rm lin}$ minimized. [This is an energy region in which
the wavefunctions tend to be scarred near $|b\rangle$ and antiscarred
near $|a\rangle$.] The relative intensity enhancement at $|b\rangle$
then scales with the height of the peak
in $S_{\rm lin}^b$, i.e. as $\lambda^{-1} \gg 1$. So a large enhancement
of the remaining probability is found on orbits out of phase with the one
on which the opening is located.

We need to consider also the case where states $|a\rangle$ and 
$|b\rangle$ are found on the same orbit [the same reasoning applies
if $|a\rangle$ and $|b\rangle$ are on distinct orbits that are related
by a symmetry transformation]. This corresponds
to measuring the remaining probability along the orbit on which the
lead is located (or alternatively to launching the initial probability
along this orbit). First, consider the case where $|a\rangle$ and
$|b\rangle$ are exactly
related by time evolution in the closed system. Then the two local
densities of states are identical, i.e.
$|\langle a|\Psi_n\rangle|^2=|\langle b|\Psi_n\rangle|^2$ for each $n$. It is
easy to see from Eq.~\ref{premb} that $P_{\rm rem}$ in this case decays at long
times as $1/t^2$ instead of the usual $1/t$ behavior. This is easy
to understand intuitively: the very long-lived resonances which survive
at long times have very little amplitude at $|b\rangle$. More generally,
let us consider $|a \rangle$ and  $|b\rangle$ lying on the same orbit but
not exact time-iterates of one another. This is possible even if
 $|a \rangle$ and  $|b\rangle$ are both optimal (in the sense of having $Q=0$,
see Eq.~\ref{alingen}). Thus, the iterates of $|a \rangle$ may have width
$\sigma_0 e^{\lambda n}$ along the unstable manifold as they pass through
that point on the orbit on which $|b\rangle$ is centered~\cite{is}.
If we choose a width for $|b\rangle$ which does not correspond to any
integer $n$,
then $|b\rangle$ is not any exact time-iterate of $|a\rangle$.
However, for some time $t$ we may still write
\begin{equation}
|b\rangle = \alpha |a(t)\rangle +\gamma |c\rangle \,,
\end{equation}
where $|\alpha|^2+|\gamma|^2=1$. Then the local density of states at $|b\rangle$
separates naturally into two parts: one of weight $|\alpha|^2$ which
is exactly equal to the density of states at the opening $|a\rangle$,
and another
of weight $1-|\alpha|^2$ which is statistically independent of the former
but has the same linear energy envelope. The first, as we just saw,
gives a contribution to $P_{\rm rem}^b$ which scales as $1/t^2$ and thus can be
ignored at long times. The second behaves just as if $|b\rangle$ were located
on a different orbit having the same linear envelope. Thus for
$|a \rangle$ and  $|b\rangle$ on the same orbit we obtain the same
exponential suppression factor as before (Eq.~\ref{ratio}), times the extra
suppression factor $1-|\alpha|^2$. This latter factor also becomes very small
for small $\lambda$ \cite{is}, as any wavepacket optimally placed on a
periodic orbit comes ever closer to being an exact time-iterate of any
other such wavepacket on the same orbit.

We note again that this effect is purely quantum-mechanical, based though
it is on short-time semiclassical analysis. Classically, only a tiny fraction
of the probability distribution corresponding to $|b\rangle$ would
leak out through the hole at $|a \rangle$ before the density escapes from the
periodic orbit and proceeds to distribute itself evenly over the entire
accessible phase space.

\section{Numerical tests}

\subsection{The model}

We now proceed to test numerically the various results obtained analytically
in the previous section. What is required is a large 
ensemble of chaotic systems with each
realization having a short unstable periodic orbit of the same instability
exponent. For this purpose we consider
kicked maps on the toroidal phase space $[-1/2,1/2] \times [-1/2,1/2]$.
The classical dynamics for one time step is given by
\begin{eqnarray}
\label{clasdyn}
p & \to & \tilde p =  p + mq - V'(q) \; {\rm mod} \;1 \\
q & \to & \tilde q = q + n\tilde p + T'(\tilde p) \; {\rm mod} \; 1\,.
\nonumber
\end{eqnarray}
This dynamics can be obtained from the stroboscopic
discretization of a kicked
system\cite{kickmap} with a kick potential $-{1 \over 2}mq^2+V(q)$ applied once
every
time step
and a free evolution governed by the kinetic term ${1 \over 2}np^2+T(p)$.
$m$ and $n$ are arbitrary
integers, while $V$, $T$ are periodic functions of position and
momentum, respectively. The system can also be thought of as a
perturbation of the linear system (cat map) \cite{pertcat}
\begin{eqnarray}
\label{catmap}
p & \to & \tilde p = p + mq \; {\rm mod} \; 1 \\
q & \to & \tilde q = np + (mn+1)q \; {\rm mod} \; 1 \,. \nonumber
\end{eqnarray}

For given positive integers $m$, $n$, we choose the functions $V$ and $T$
such that $m-V''(q)>0$ for all $q$ and similarly
$n+T''(\tilde p)>0$ for all $\tilde p$. Then the system is strictly
chaotic and looks everywhere locally like an inverted harmonic oscillator.

The quantization of such systems is well-studied in the literature
\cite{kickmap}. $\hbar$ must be chosen such that $N=1/2\pi\hbar$,
the number of $h-$sized cells in the classical phase space, is an
integer ($N$ must be even to preserve the periodicity of the
quadratic terms in the potential and kinetic energy).
For doubly-periodic boundary conditions,
the quantum $N-$dimensional Hilbert space is spanned
by the position basis $|q_i\rangle$, where $q_i=i/N$
and $i=0 \ldots N-1$. The momentum-space basis is given similarly
by $|p_j\rangle$, where $p_j=j/N$ and
$j=0 \ldots N-1$; and the two bases are related by a discrete
fourier transform. The quantum dynamics is then given by a unitary
$N \times N$ matrix
\begin{eqnarray}
U & = &
\exp{\left[-i \left({1 \over 2} n\hat p^2 +
T(\hat p)\right)/\hbar\right]}
\nonumber \\ & \cdot &
\exp{\left[i \left({1 \over 2} m\hat q^2 -V(\hat q)\right)/\hbar\right]}
\,.
\end{eqnarray}
Each factor is evaluated in the appropriate basis, and an implicit
forward and backward fourier transform has been performed.

We may now perturb this unitary dynamics to allow for a small
decay rate in channel $a$:
\begin{equation}
\label{wdef}
W=\left(1-{\Gamma \over 2} |a\rangle\langle a|\right)U \,.
\end{equation}
Eq.~\ref{wdef} is of course the discrete-time version of the continuous-time
dynamics given in Eq.~\ref{hdef} above. Since we are working in the regime
$\Gamma \ll 1$, where the decay rate per time step is small, the
discretization of the decay process will not affect the long-time behavior
of the system. The decay channel $|a\rangle$ can in principle 
represent any vector in the Hilbert space; however, for the decay to
correspond to a classical escape route, $|a\rangle$ should be a phase-space
localized state, such as a position or momentum state. We will find it
convenient to let $| a\rangle$ be a circular phase-space gaussian
in the coordinates $q$, $p$.

We now need to construct an ensemble of such systems, all having the same
behavior in the vicinity of a short unstable periodic orbit. For this purpose,
we set $m=n=1$, and let the potential $V$ and kinetic term $T$ be odd
functions of their respective arguments:
\begin{eqnarray}
\label{vt}
V(q) & = & \sum_{r=1}^3 [K_r \sin(2\pi rq) - 2\pi K_rrq]  \\
T(p) & = & \sum_{r=1}^3 [K'_r \sin(2\pi rp) - 2\pi K'_rrp] \,.
\end{eqnarray}
It is easy to see that the equations of motion (Eq.~\ref{clasdyn})
then have a fixed point at the origin with Jacobian matrix  
$J =\left[\begin{array}{cc} 1 & 1 \\ 1 & 2 \end{array}\right]$ [the system
can be thought of as a perturbation of the 
$\left[\begin{array}{cc} 1 & 1 \\ 1 & 2 \end{array}\right]$ cat map, with
perturbation vanishing near $(q,p)=(0,0)$]. The instability exponent 
is given by $\lambda= \cosh^{-1} \left({1 \over 2} {\rm Tr} J \right) = 0.96$.
We may now
choose each of
the coefficients $K_r$, $K'_r$ from a uniform distribution 
over the interval $\left[-{0.3 \over (2\pi r)^2},{0.3 \over (2\pi r)^2}\right]$.
One easily
sees that each system in this ensemble satisfies everywhere the condition 
$1-V''>0$, $1+T''>0$ mentioned above, which is sufficient to ensure hard
chaos.

\subsection{Probability to remain}
\label{numprob}

The ensemble-averaged probability to remain in the system after $t$ time
steps is now computed for various positions of the exit channel $|a\rangle$.
For simplicity, we choose $|a\rangle$ to be a circular phase-space gaussian
\begin{equation}
a(q) \sim e^{-(q-q_0)^2/2\hbar+i p_0(q-q_0)/\hbar}
\end{equation}
centered on $(q_0,p_0)$ and having width  $\sqrt\hbar$ in both the $q$
and $p$ directions. Because the Jacobian $J$ at the fixed point $(0,0)$
is symmetric (and the stable and unstable directions are thus locally
orthogonal), such a wavepacket, when centered at $(0,0)$, is optimal
in the sense of having the slowly decaying
short-time autocorrelation function of Eq.~\ref{alin}. [A more general 
gaussian wavepacket (including position or momentum states as extreme limits)
centered on the periodic orbit may have a non-zero parameter $Q$ (see
Eq.~\ref{alingen}), leading to a less sharp linear spectral
envelope and less strong scarring and antiscarring effects. The qualitative
behavior would, however, remain unchanged.]

In Fig.~\ref{fig1}, the probability to remain in the system as a function
of the scaled time $t'=\Gamma_{\rm cl}t$ is first plotted (using plusses)
for a generic
leak location. The data was collected for systems of size 
$N=120$ and decay parameter $\Gamma=0.1$. The results agree well with the
the RMT prediction $P_{\rm rem}(t')=1/(1+t')$ (dashed curve). For comparison,
the classical probability to remain, $\exp(-t')$, is plotted as a dotted
curve. Next, we place the opening on the periodic orbit at the origin
of phase space, and obtain the rather different behavior, with an enhanced
long-time tail (squares). The asymptotic form is well reproduced by the
scar theory prediction, $P_{\rm rem}(t') \approx <S_{\rm lin}^{-1}> /
t'$, which is shown in Fig.~\ref{fig1} as a solid line. For the instability
exponent $\lambda=0.96$, we observe a long-time probability enhancement factor
$<S_{\rm lin}^{-1}>=11.04$. Of course, bigger enhancement factors can be
observed for less unstable orbits, as we will see below. 

First, we examine more carefully the probability to remain at long times
as a function of the position of the lead. In Fig.~\ref{fig2}a is plotted the
total probability to remain in the system at time $t'=2\cdot 10^3$, for various 
locations of the opening (all for $N=30$).
These possible locations are located on
a $40 \times 40$ grid filling the middle $1/9$th section of the total
phase space (i.e. $(q,p)\in [-1/6,1/6] \times [-1/6,1/6]$). The bright spot
at the center of the figure represents the enhanced probability to remain if the
opening is located exactly on the periodic orbit. As we see from the figure,
the antiscarring effect falls off quite quickly as the opening is moved away
from the periodic orbit (in fact, the size of the bright spot scales
as $\hbar$; see Eq.~\ref{deltaeq} and discussion following).
In Fig.~\ref{fig2}b is plotted the theoretical quantity
$<S_{\rm lin}^{-1}>$, as computed using the linearized equations of motion
(Eq.~\ref{catmap}) around the periodic orbit. This is observed to be in good
agreement with the data. Of course the linearized equations of motion only
hold near the periodic orbit itself, and do not correctly describe
classical motion in other regions of phase space. However, in our case
the short periodic orbit at the origin clearly dominates the data. If
the classical system contained several not very unstable orbits (see next
subsection
for an example), several bright spots would appear in the plot, and each could
then be well reproduced using the linearized classical dynamics around the
appropriate orbit.

An important feature to notice in Fig.~\ref{fig2} is that $P_{\rm rem}(t)$
at long times depends not only on the distance of the lead from the periodic
orbit but also on the direction. Greater enhancement is observed if the lead
is placed along either 
the stable or the unstable manifold of the orbit (the two ``diagonals"
of the would-be ``square").

We now consider how the probability to remain at long times depends on the
instability exponent $\lambda$ of the orbit on which the lead is located. 
As we discussed in the previous section, as $\lambda$ gets small, 
resonances exponentially narrow in $\lambda$ should appear at the
antiscarring energies, and the total probability to remain at long times
is enhanced by a factor exponentially large in $\lambda$ (Eq.~\ref{enh}).
To see this effect, we modify the classical dynamics of Eq.~\ref{clasdyn}
by adding an additional term to the potential and kinetic functions:
\begin{eqnarray}
V(q)&=& \ldots - {\Delta \over (2\pi)^2} \cos(2\pi q) \nonumber \\
T(p)&=& \ldots + {\Delta \over (2\pi)^2} \cos(2\pi p) \,.
\end{eqnarray}
The same value of $\Delta$ should be used in the potential and kinetic
terms to preserve the symmetry of the Jacobian. The Jacobian matrix of the
dynamics near the periodic orbit at $(0,0)$ is then given by
\begin{equation}
\left[\begin{array}{cc} 1 \; & 1-\Delta \\ 1-\Delta \; & 
1+(1-\Delta)^2 \end{array}\right]
\end{equation}
For positive $\Delta$, the trace of the Jacobian decreases and the
orbit becomes less unstable. Using $\Delta=0.0$, $0.1$, $0.2$, $0.3$, and
$0.4$, we obtain exponents $0.96$, $0.87$, $0.78$, $0.68$, and $0.59$,
respectively.

In Fig.~\ref{fig3}
the long-time enhancement factor of $P_{\rm rem}(t)$ over its RMT value
is plotted as a function of the exponent $\lambda$,
using plusses for $N=120$ and squares for $N=240$. The theoretical
prediction $<S_{\rm lin}^{-1}>$ is shown as a solid curve. The data
consistently falls below the theoretical prediction, with the
disagreement becoming more pronounced at the smaller
values of $\lambda$. The reason is primarily a finite-size effect:
the analytical calculations are all carried out under
the assumption that the mean level spacing is much smaller than any scale
over which the linear energy envelope changes significantly. Then the linear
envelope is roughly constant on the scale at which individual resonances
emerge, and their behavior can be treated statistically. Thus, the discrepancy
becomes more noticeable as $\lambda \to 0$ for fixed $N$, as the structures
in $S_{\rm lin}$ become more comparable to the mean level spacing. Indeed,
we see that the $N=240$ data is consistently closer than the
$N=120$ data to the theory, which
strictly applies only in the semiclassical $N \to \infty$ limit.

We observe the exponential increase in the enhancement factor as
$\lambda$ decreases; indeed the very moderate exponent $\lambda=0.59$
produces
an enhancement factor of well over $100$ in the long-time
probability to remain.
$\lambda=0.1$ would in theory produce an average long-time enhancement of
$1.2 \cdot 10^{13}$, provided we were able to go to a large enough
system, wait for a long enough time, and collect enough statistics to
observe it [from Eq.~\ref{preme} we see that the asymptotic
$1/S_{\rm lin}t'$ form holds only for $t' \gg S_{\rm lin}^{-1}$].

\subsection{Phase space distribution of long-time probability density}

We now consider predictions concerning the enhancement or
suppression of the long-time probability near orbits {\it other} than
the one on which the lead is located [see discussion in Section~\ref{dens}].
To test these predictions we need an ensemble of systems all having two
periodic orbits in common, and the ability to vary the action phase
difference between them. In our example the two orbits have the same local
dynamics in their respective neighborhoods, though this of course is not
necessary to produce the desired effect.

We work on the phase space $(q,p) \in [-1/4,3/4] \times [-1/2,1/2]$,
set $m=n=2$ in the equations of motion (Eq.~\ref{clasdyn}), and
impose the constraint $K_1=-3 K_3$ on the kick potential (see Eq.~\ref{vt}).
This condition ensures the presence of a fixed point at (1/2,0)
in addition to the usual one at (0,0) on which we have been focusing so
far.
The linearized dynamics around each orbit is given by the Jacobian
$J =\left[\begin{array}{cc} 1 & 2 \\ 2 & 5 \end{array}\right]$, and
the exponent per period is
$\lambda=1.76$. Other orbits are of course present, but they change with
the coefficients $K_r$, $K'_r$, and so their effects are expected to 
cancel out in the process of ensemble averaging. In order for the two orbits not
to be related by a symmetry transformation we only need all the $K_r$
to be nonvanishing. 

Our analysis showed that the behavior of the remaining probability
density at long times should depend strongly on the relative action
phase difference between the two orbits. This phase difference can
be easily controlled by adjusting $K_2$ ($\phi_b - \phi_a=N K_2/4$).
Fixing $K_2$ at a non-zero value which produces $\phi_b - \phi_a=0
\,({\rm mod}\, 2\pi)$, we are free to vary $K_3$ and the three coefficients
$K_r'$ consistent with the constraints $2-V''>0$ and $2+T''>0$ (which
are sufficient to ensure hard chaos). We obtain then,
for $N=80$ and $\Gamma=0.1$, the
results shown in Fig.~\ref{figrem}a. Clearly the remaining probability
is very strongly suppressed on the orbit on the left, where the lead
is located [as we saw in the previous ection, the probability to remain
{\it exactly} on the orbit falls off faster with time than probability 
elsewhere, so the numerical value of the relative suppression
there will be time-dependent]. We also see mild density suppression on the two
orthogonal invariant manifolds of this orbit. The phenomenon we want to focus
on here, though, is the suppression we observe on the orbit on the right
side of Fig.~\ref{figrem}a. The observed suppression factor right on the
periodic orbit at $(1/2,0)$ is
$0.37$, compared with the predicted value $0.45$; again the discrepancy may
possibly be attributed to finite size effects. As expected, we also
observe probability suppression along the manifolds of this second orbit.

We now consider the opposite case, where the two orbits are exactly out 
of phase ($\phi_b - \phi_a=\pi \,({\rm mod}\, 2\pi)$).
Adjusting $K_2$ appropriately, we again perform ensemble averaging over the
other parameters and obtain the results in Fig.~\ref{figrem}b. The same
suppression is still seen along the orbit containing the lead and its
invariant manifolds. However, we now see, as expected, an {\it enhancement} of 
the remaining probability density near the orbit on the right. The enhancement
factor on that orbit itself is $1.78$; the theoretical prediction
is $1.67$. 

The predicted relative intensity at the orbit $(1/2,0)$ (Eq.~\ref{ratio})
given a lead
at $(0,0)$ is plotted in Fig.~\ref{fig2po} as a function of the
action phase difference between the two orbits. [One could imagine obtaining
such a plot in a physical system by tuning a weak magnetic field which
had little effect on the classical dynamics but did change the relative
phase between two orbits enclosing different amounts of flux. Alternatively,
if the periods of the two orbits differed, the orbits could be observed
to go in and out of phase with one another as one changed the energy range
in which the resonances were populated.]
The suppression and
enhancement factors in this case never get very far from unity,
due to the relatively large value of the instability exponent $\lambda$
chosen for our example. Values obtained numerically, as described in
the preceding paragraphs, are given for comparison, along with statistical error
bars. The dashed line at $1$ represents the RMT prediction.

\section{Conclusion}

We have concentrated throughout on leakage through a single decay channel
which is very well localized in phase space (leak area much less than $h$). One
would like to understand more generally possible non-RMT effects for
multi-channel leads as well as for leads wide enough to produce
overlapping resonances. A detailed treatment of such effects is outside
the scope of the present work. However, one possible
extension turns out to be relatively straightforward,
and the results suggest that the most interesting scar effects are already
captured in the single channel analysis. Specifically, consider $M$
decay channels, each with a slow decay rate, as in Eq.~\ref{hmany}. If
the sum of these classical rates is small compared to the level spacing,
the resulting resonances will be non-overlapping, and a perturbative approach
to the problem is valid. In RMT, the probability to remain at long times
is given by a product of factors associated with each of the leads
(Eq.~\ref{pmany}). In the case where one of the channels happens to
be close to a short periodic orbit, the intensity distribution giving
rise to the corresponding factor only will be affected, producing the
same overall enhancement factor $<S_{\rm lin}^{-1}>$ obtained previously.
The total probability $P_{\rm rem}$ now falls off much more quickly
with time than in the single-channel case, reflecting the fact that it
is much harder to find a resonance which is slowly decaying through
{\it all} of the channels. However, the essential
observation of an enhanced
probability of finding very narrow resonances survives. Clearly
the analysis is more complicated for correlated decay channels, and
for the case of larger total decay rates where the resonances become
overlapping. Still, periodic
orbits by their very nature produce quantum effects in phase space regions
of size $\hbar$ surrounding the orbit; thus our intuition tells us that
no fundamentally new scar effects are expected in most cases for multi-channel
leads.

We have seen that an analysis of short time classical motion in a chaotic
system can shed much light on quantum behavior on the scale of the decay time,
which is
much larger than the Heisenberg time and every other time scale in the system.
This is somewhat counterintuitive, as the narrow resonance regime
is by its very nature non-classical and is associated with the very long-time
behavior of the system. Even though in some cases (e.g.
in the presence of strong diffraction or caustics)
semiclassical methods
may not be sufficient to predict the properties of individual high-energy
quantum chaotic 
wavefunctions, they are still very powerful for making statistical predictions
of the sort described in this paper. We note that a small change in the
system potential, or the presence of a few impurity scatterers may
completely change the character of individual resonances in an open
system, making comparison with exact semiclassical wavefunctions futile.
On the other hand, the scar theory predictions, which concern
statistical properties such as the distribution of resonance widths,
are robust to such changes in the details of the system, as long as the
dynamics of the first few bounces is known. In a situation where
the classical dynamics of the quantum system under study is not known
reliably even for short times, one could use methods similar to those
described here to search for the short unstable periodic orbits. This can
be done by moving the position of the lead, or, more practically
in many situations, by adjusting some system parameter which changes
the classical dynamics of the system. For a given classical system and lead
position, one then sweeps through a weak magnetic field or some 
other parameter not affecting the classical dynamics, and searches for
a large fraction of very narrow resonances,
occurring periodically in the
magnetic field strength.

Clearly the ideas described here can also be extended to study two-lead
systems, where properties such as conductance peak height
distributions can be analyzed. In analogy with the present work, the results
will strongly depend on whether one or both of the leads is located on
a short unstable classical orbit. Where the leads are found on
two different short periodic orbits, the phase difference between them can be
varied to produce an enhancement or suppression of the average
conductance, as suggested by the results of the present work. Even stronger
effects can be observed if the leads are located on the same periodic orbit,
or if the two orbits
are related to each other by a symmetry of the system. These
issues and other related questions are addressed fully in a forthcoming
paper~\cite{bies}. Certainly much more work needs to be done generally in
understanding classical dynamics effects on the quantum properties 
of open chaotic systems.

\section{Acknowledgements}

This research was supported by the National Science Foundation under
Grant No. 66-701-7557-2-30. Initial work on this project was performed
during a stay at the Technion in Israel. The author thanks E. J. Heller
for many useful conversations.

\begin{figure}
\caption{Smoothed local densities of states $S_{\rm lin}(E)$ are plotted as
a function of energy on a periodic orbit of instability exponent $\lambda=0.20$
(solid curve) and on an orbit with $\lambda=0.15$ (dashed curve). The mean
resonance width for a lead placed on such a periodic orbit will be
proportional to $S_{\rm lin}(E)$. We observe
the peak at the EBK quantization energy $E=0$ (Eq.~\ref{maxs}) which scales
as $\lambda^{-1}$, the exponential decay between $E=0$ and $E=\pi$
(Eq.~\ref{inters}), and the minimum at the anti-EBK energy $E=\pi$,
which is exponentially small in $\lambda$
(Eq.~\ref{mins}). The RMT prediction
$S_{\rm lin}(E)=1$, which is applicable away from any short periodic orbit,
is plotted as a dotted line.
}
\label{figslin}
\end{figure}

\begin{figure}
\caption{The probability to remain in the open quantum system is
plotted as a function of scaled time $t'=\Gamma_{\rm cl}t$. The classical
prediction $\exp (-t')$ is shown as a dotted curve. The quantum probability
to remain for a generic lead location (plusses) compares well with the
RMT prediction $1/(1+t')$ (dashed curve). For a lead placed on a short
periodic orbit with instability exponent $\lambda=0.96$, we obtain the enhanced
long-time probability to remain (squares), which agrees with the scar
theory prediction $<S_{\rm lin}^{-1}>/t'$ (solid line). The system size used
for obtaining the data is $N=120$ and the decay rate per step
in the exit channel is $\Gamma=0.1$.
}
\label{fig1}
\end{figure}

\begin{figure}
\caption{The remaining probability density after time $t'=2 \cdot 10^{-3}$,
as a function of the position of the lead. In (a), the numerical data is
presented for an ensemble of systems of size $N=30$; in (b) we show
the theoretical prediction $<S_{\rm lin}^{-1}>/t'$. At the center of each
plot is an unstable periodic orbit of exponent $\lambda=0.96$: for a lead
placed at that position (white spot) the probability at long times is enhanced
by a factor of $11$ over the same probability for a generic lead position
(black background). Notice the anisotropy: more enhancement at long
times is predicted (and observed) when the displacement of the lead
away from the periodic orbit is along one of the invariant manifolds.
}
\label{fig2}
\end{figure}

\begin{figure}
\caption{The long-time enhancement factor of the probability to remain
in a system when the lead is placed on a periodic orbit is plotted
as a function of the instability exponent of the orbit. Data is shown
for $N=120$ (plusses) and $N=240$ (squares).
The $N \to \infty$ theoretical
prediction $<S_{\rm lin}^{-1}>$ is shown as a solid curve. We see the
exponential increase in the probability to remain as the exponent $\lambda$
decreases (the $\lambda \to 0$ asymptotic form is given in Eq.~\ref{enh}).
For large $\lambda$, the enhancement factor converges to $1$, the RMT
prediction.
}
\label{fig3}
\end{figure}

\begin{figure}
\caption{The remaining probability distribution at very long times is 
shown for an ensemble of systems (of size $N=80$)
all having two short periodic orbits in
common, each with instability exponent $\lambda=1.76$. In both cases, the
lead is centered on the periodic orbit on the left side of the plot. The
remaining probability is strongly suppressed on that orbit and less so
on its invariant manifolds. The action of the
orbit on the right is chosen to be in phase with the first one in case (a),
so that probability there is also suppressed, and exactly out of phase with
it in case (b), leading to an enhancement of the probability density on the
second orbit and its invariant manifolds.
See next figure for quantitative
comparison with the theory.
}
\label{figrem}
\end{figure}

\begin{figure}
\caption{
The predicted relative intensity at long times on a periodic orbit other
than the one containing the lead is plotted as a function of the relative
phase between the two orbits (solid curve). For reference, the phase-space
averaged intensity is plotted as a dashed line. Intensity suppression
is predicted and observed when the orbits are in phase, and enhancement
is seen when the orbits are exactly out of phase [compare (a) and (b)
in previous figure]. The data is collected for an ensemble of systems with
$N=80$, and each orbit has instability exponent $\lambda=1.76$.
}
\label{fig2po}
\end{figure}

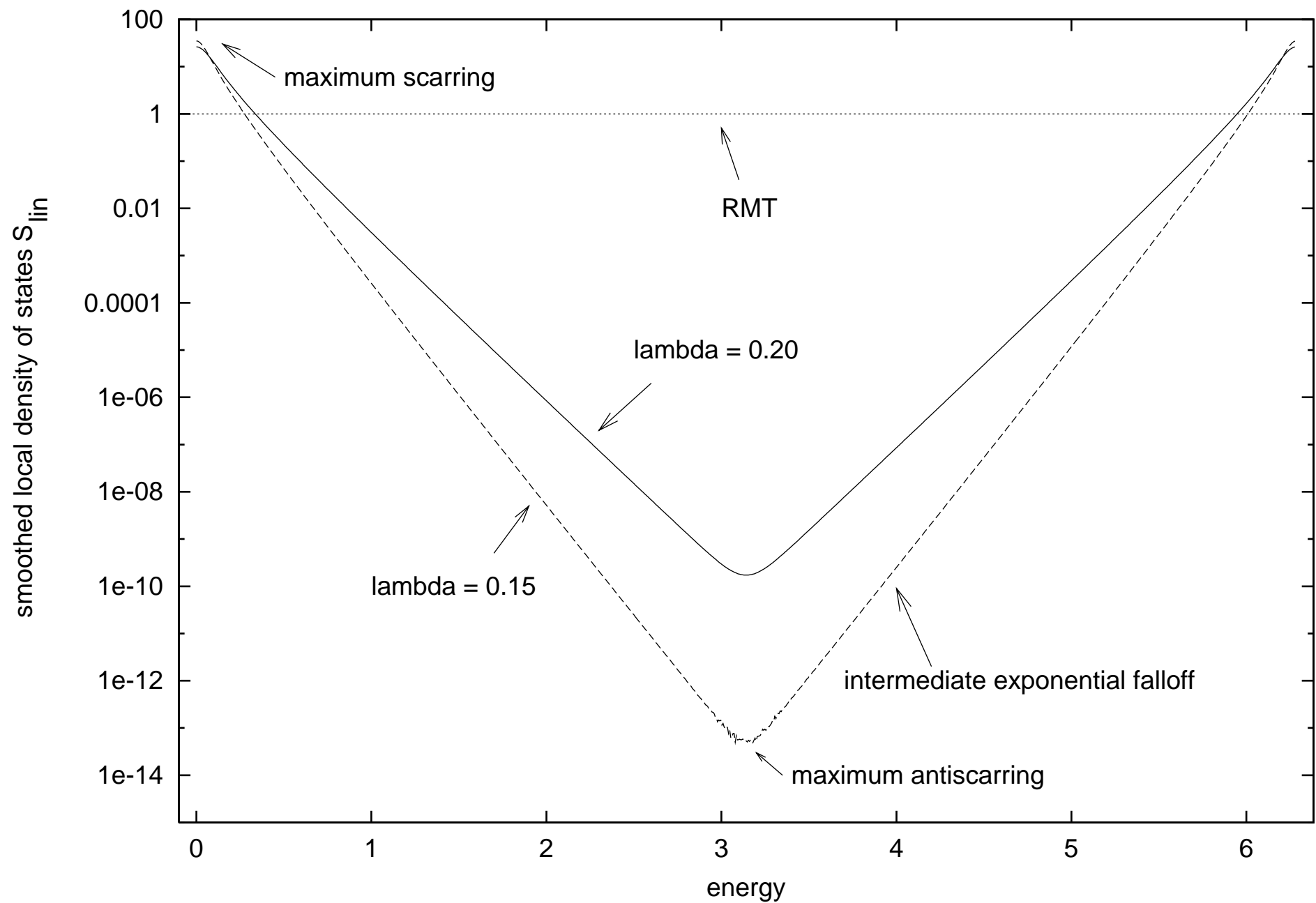
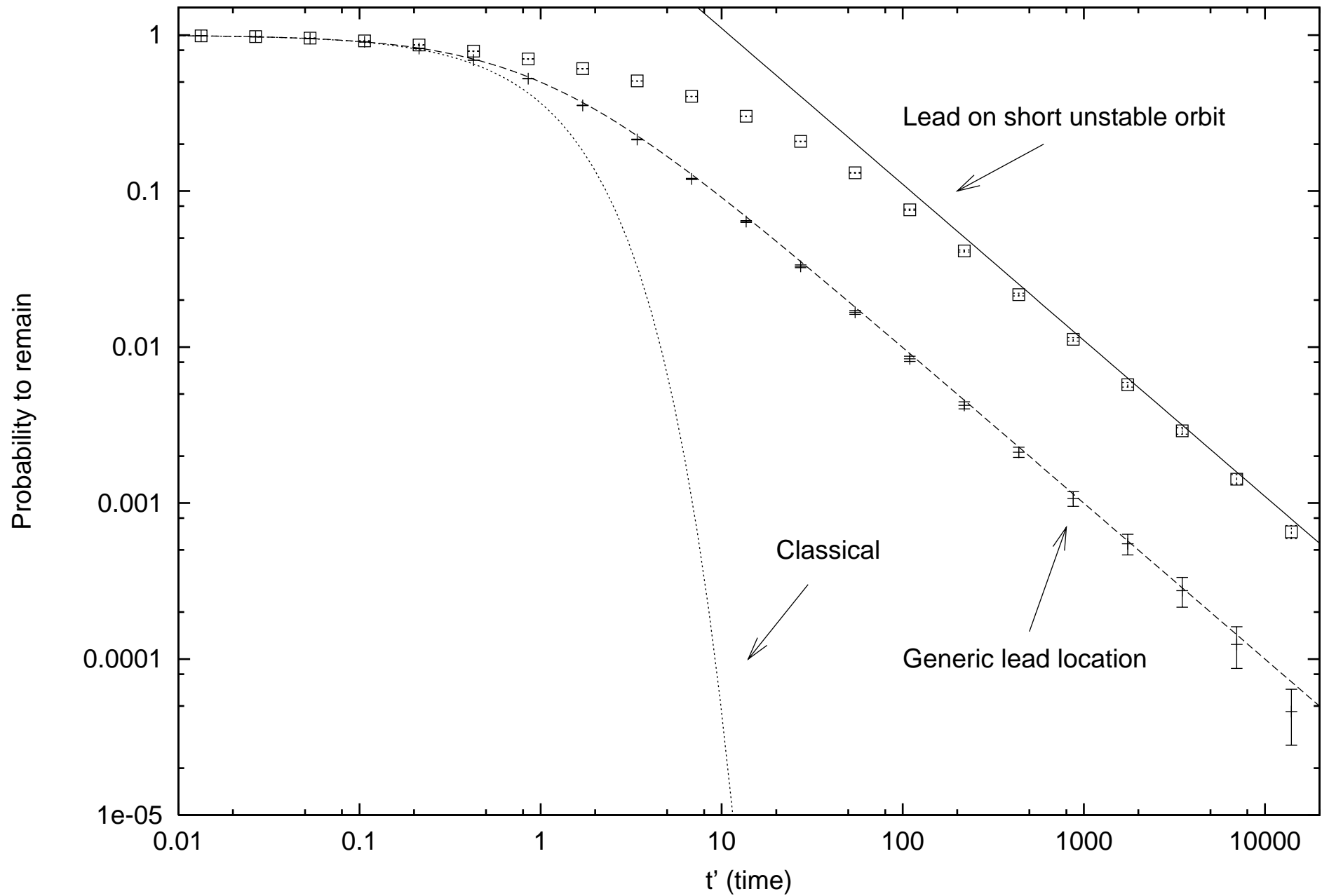
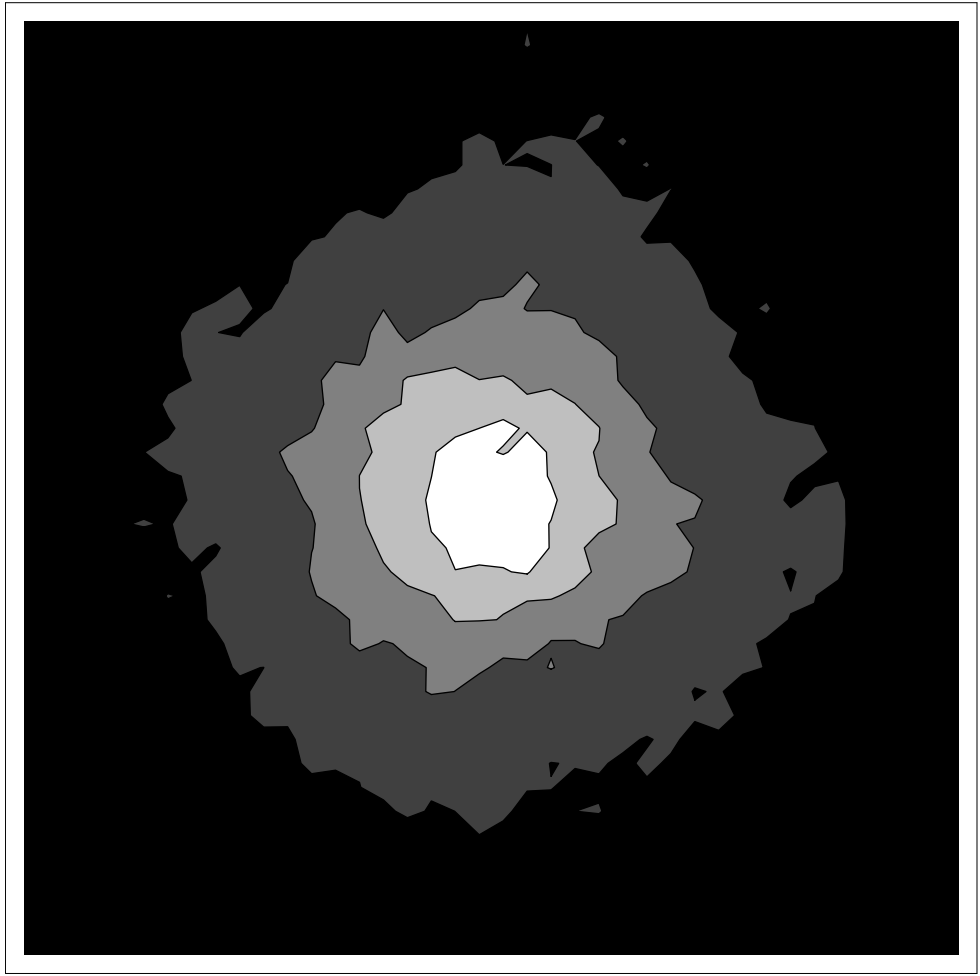
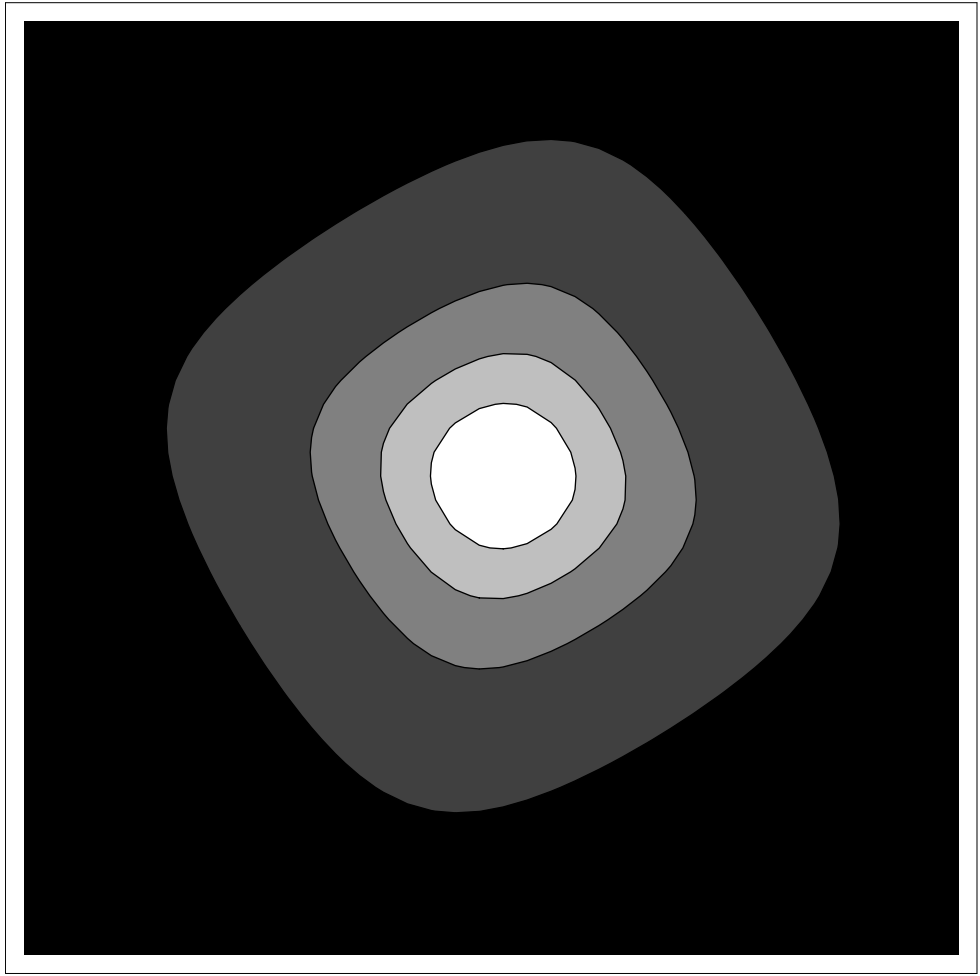
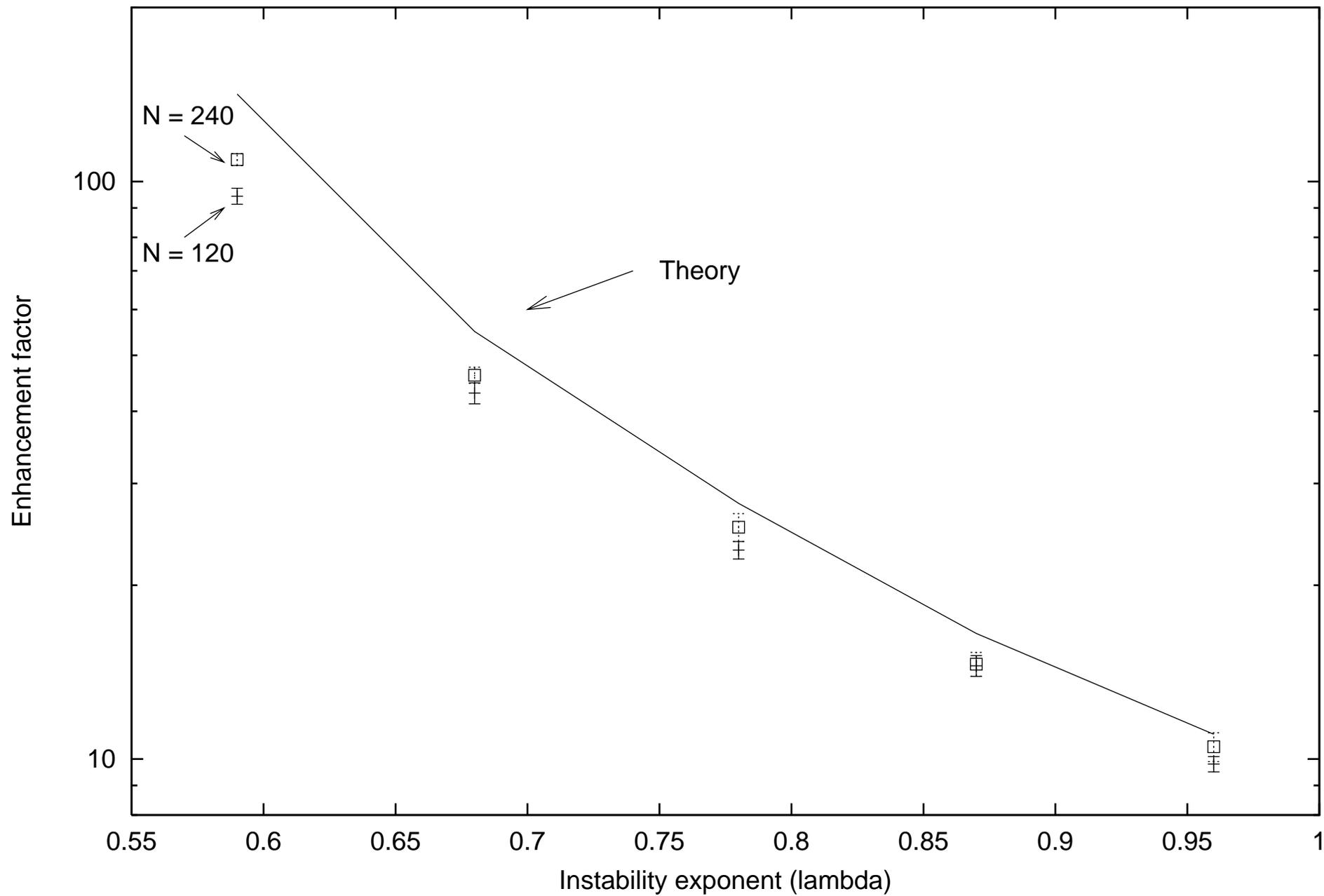
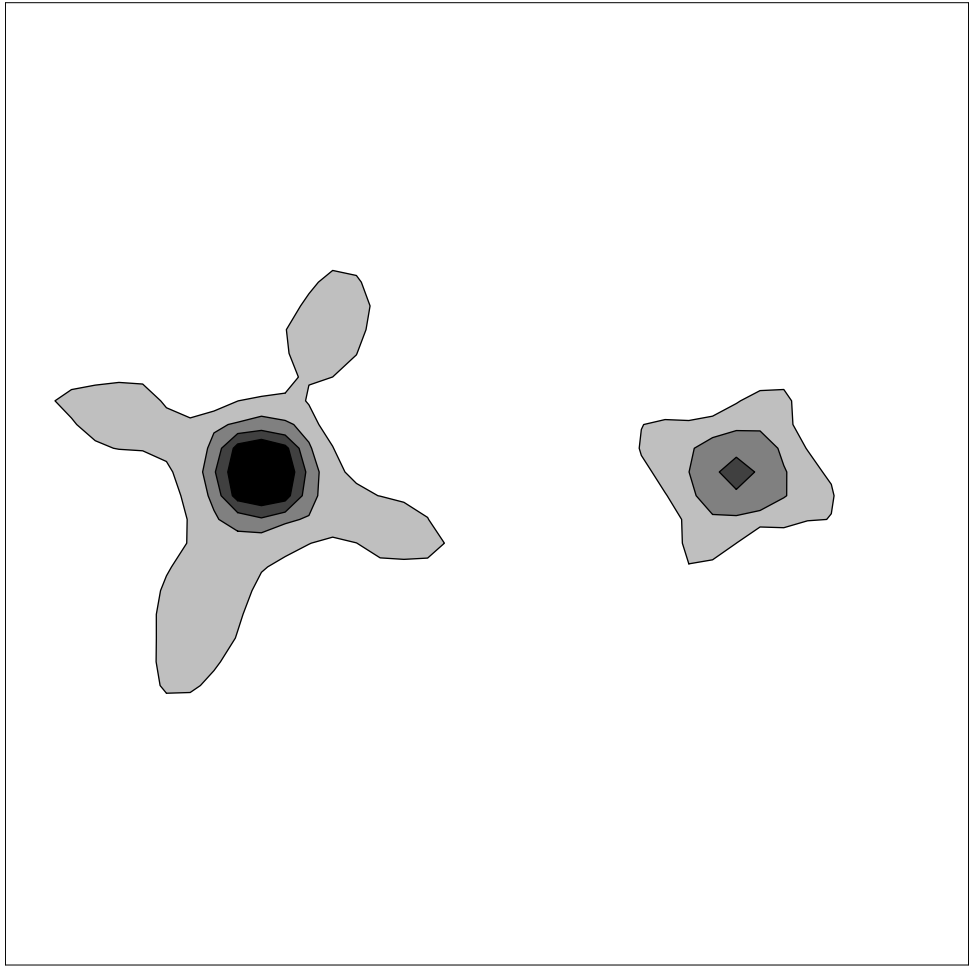
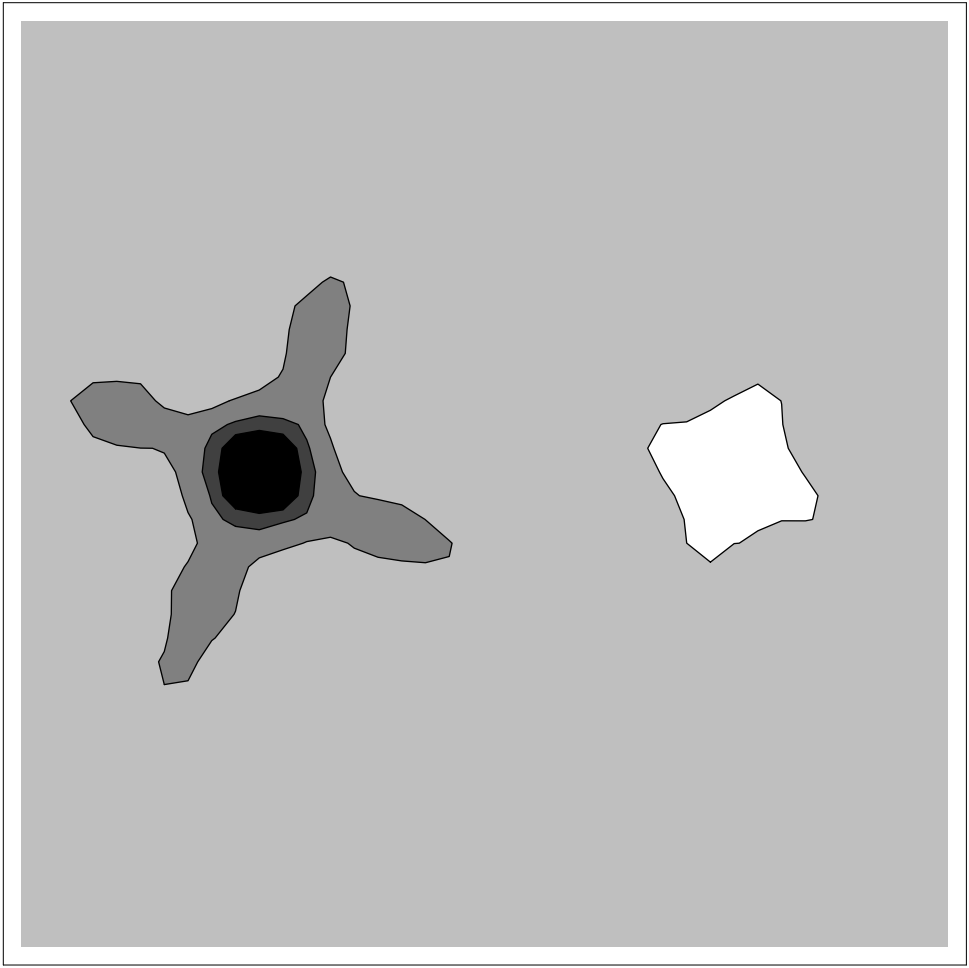
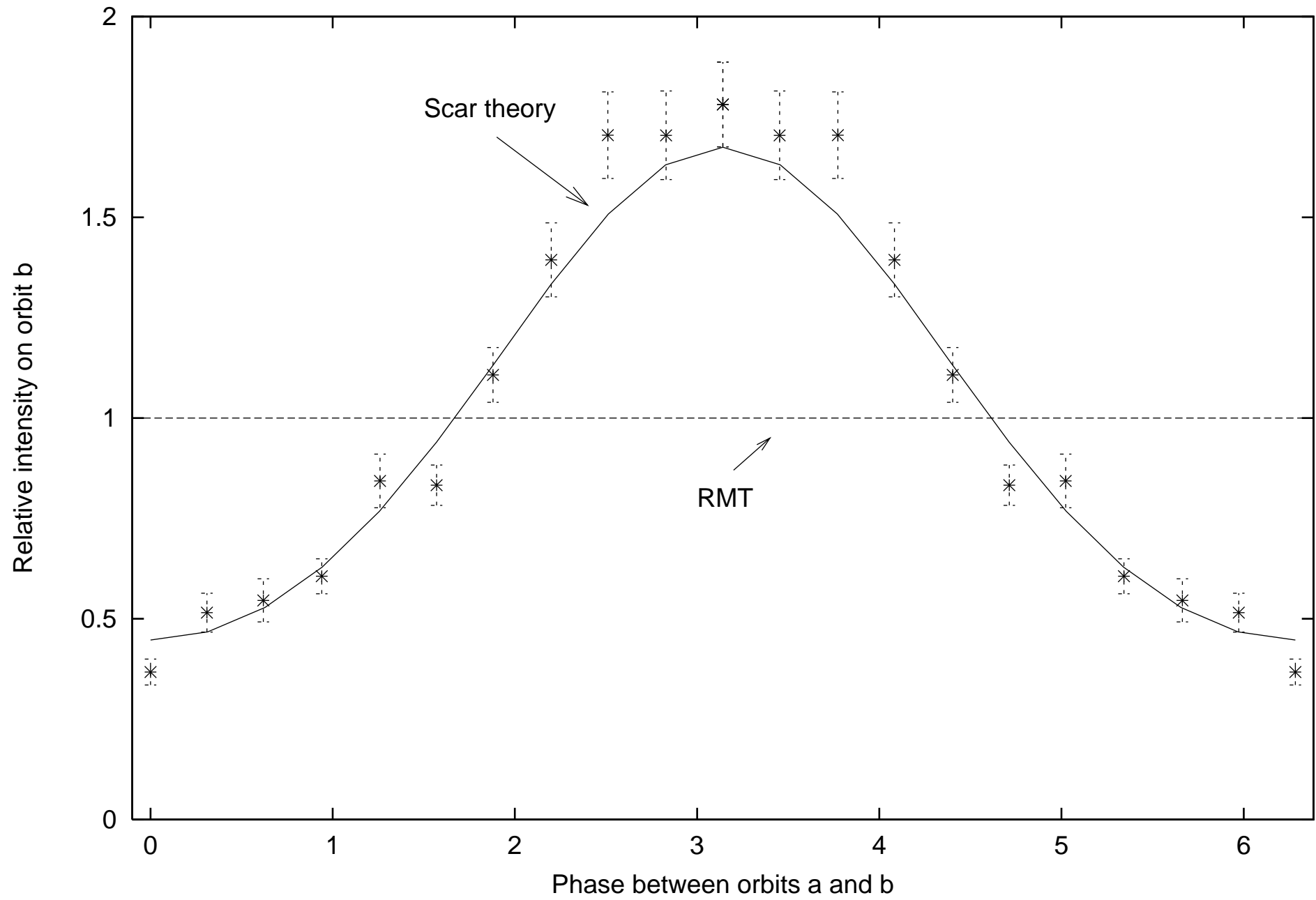
\end{document}